\documentstyle[prl,preprint,aps]{revtex}

\begin{document}
\bibliographystyle{prsty}
\title{Sub-natural linewidth in room-temperature Rb vapor 
using a control laser}
\author{Umakant D. Rapol, Ajay Wasan, and Vasant 
Natarajan\thanks{Electronic address: 
vasant@physics.iisc.ernet.in}}
\address{Department of Physics, Indian Institute of 
Science, 
Bangalore 560 012, INDIA}

\maketitle
\begin{abstract}
We demonstrate two ways of obtaining sub-natural 
linewidth for probe absorption through room-temperature 
Rb vapor. Both techniques use a control laser that drives 
the transition from a different ground state. The 
coherent drive splits the excited state into two dressed 
states (Autler-Townes doublet), which have asymmetric 
linewidths when the control laser is detuned from 
resonance. In the first technique, the laser has a large 
detuning of 1.18 GHz to reduce the linewidth to 5.1 MHz 
from the Doppler width of 560 MHz. In the second 
technique, we use a counter-propagating pump beam to 
eliminate the first-order Doppler effect. The unperturbed 
probe linewidth is about 13 MHz, which is reduced below 3 
MHz ($0.5 \Gamma$) at a detuning of 11.5 MHz.
\end{abstract}
\pacs{42.50.Gy,42.50.Md,42.50.Vk}

Laser spectroscopy in a room-temperature gas is often 
limited by Doppler broadening due to the thermal velocity 
of gas particles. However, techniques such as 
saturated-absorption spectroscopy using counter-propagating pump 
and probe beams can be used to eliminate the first-order 
Doppler effect and linewidths close to the natural 
linewidth can be obtained. But overcoming the natural 
linewidth is not easy and it appears as a fundamental 
limit to the resolution that can be achieved in precision 
spectroscopy. The natural linewidth also plays a role in 
other areas. For example, in laser cooling of atoms and 
ions, the lowest attainable temperature (Doppler limit) 
is determined by the natural linewidth of the cooling 
transition. In addition, when lasers are locked to atomic 
transitions, the natural linewidth determines the 
tightness of the lock. It is therefore desirable to 
develop techniques for getting below the natural 
linewidth.

In this Letter, we demonstrate two techniques to obtain 
sub-natural linewidth in a room-temperature vapor of Rb 
atoms. The techniques have been adapted from recent 
developments in the use of control lasers in three-level 
systems as a means of modifying the absorption properties 
of a weak probe beam \cite{NSO90}. For example, in 
electromagnetically induced transparency (EIT), an 
initially absorbing medium is made transparent to the 
probe when a strong control laser is turned on 
\cite{BIH91}. We use a $\Lambda$-type system, where the 
control laser drives the transition from one ground 
state, and the probe laser measures absorption from the 
other ground state. The control laser creates two dressed 
states \cite{COR77} and the probe absorption splits into 
an Autler-Townes doublet. The key idea is that, when the 
control laser is detuned from resonance, the two dressed 
states have asymmetric linewidths, {\it such that their 
sum is the unperturbed linewidth} \cite{VAR96,AGA96}. 
This asymmetry persists even when the peaks are 
inhomogeneously broadened. Thus, for suitable detuning, 
the width of one state can be much smaller than the 
unperturbed linewidth.

There have been previous experimental studies on 
linewidth narrowing using a control laser, but this was 
in the spontaneous-emission spectrum of a three-level V 
system \cite{GZM91}. The experiments were done using an 
atomic beam where the Doppler broadening was negligible. 
By contrast, our experiments show 
sub-natural linewidth in the {\it absorption spectrum} 
and are done in 
room-temperature vapor with Doppler broadening of 560 
MHz. Furthermore, we use a $\Lambda$ system and the basic 
mechanism for linewidth narrowing is different. In the V 
system, the ground state is coupled by the control laser 
to a weak auxiliary transition. Spontaneous emission from 
the excited state is suppressed because the quantum 
fluctuations in the {\it ground state} are stabilized by 
the control laser, as shown by Narducci {\it et al.\ 
}\cite{NSO90}. In our case, linewidth narrowing occurs 
because of quantum coherences in the {\it excited state} 
created by the control laser.

To understand these ideas more clearly, let us consider 
the three-level $\Lambda$ system in $^{87}$Rb in greater 
detail. The relevant energy levels are shown in Fig.\ 
\ref{levels}. The lower levels $\left| 1 \right>$ and 
$\left| 2 \right>$ are the $F=1$ and $F=2$ hyperfine 
levels in the $5S_{1/2}$ ground state, while the upper 
level $\left| 3 \right>$ is the $5P_{3/2}$ excited state. 
The excited state has four hyperfine levels; of these, 
only the levels with $F'=1$ and 2 couple to both the 
lower levels and can act as level $\left| 3 \right>$. The 
control laser drives the $\left| 1 \right> 
\leftrightarrow \left| 3 \right>$ transition with Rabi 
frequency $\Omega_R$ and detuning $\Delta_c$. The weak 
probe laser measures the absorption on the $\left| 1 
\right> \rightarrow \left| 2 \right>$ transition at a 
detuning $\Delta$. The excited state lifetime is 26 ns, 
threfore the spontaneous decay rates $\Gamma_{31}$ and 
$\Gamma_{32}$ are both equal to $2\pi \times 6.1$ MHz.

The absorption of the weak probe in the $\Lambda$ system 
has been derived previously \cite{VAR96}. As is well 
known, the control laser splits the upper level into two 
dressed states due to the ac Stark shift. The probe 
absorption gets modified due to this and shows peaks at 
the location of the two dressed states (Autler-Townes 
doublet), given by 
\begin{equation}
\Delta_{\pm} = \frac{\Delta_c}{2} \pm 
\frac{1}{2}\sqrt{\Delta_c^2 + \Omega_R^2}.
\end{equation}
Here $\Delta_{+}$ and $\Delta_{-}$ are the values of the 
probe detuning where the peaks occur. The corresponding 
linewidths ($\Gamma_{\pm}$) of these peaks are different 
because of the coherence between the two dressed states, 
and given by
\begin{equation}
\Gamma_{\pm} = \frac{\Gamma_{31} + \Gamma_{32}}{4} 
\left( 1 \mp \frac{\Delta_c}{\sqrt{\Delta_c^2 + 
\Omega_R^2}} \right) .
\end{equation}
It is clear from the above expression that, if $\Delta_c 
= 0$, the two peaks are symmetric and have identical 
linewidths of $\left( \Gamma_{31} + \Gamma_{32} \right) 
/4$. However, for any non-zero detuning, the peaks have 
asymmetric linewidths. The first peak has larger 
linewidth while the second peak has smaller linewidth by 
precisely the same factor, in such a way that the sum of 
the two linewidths is equal to the unperturbed linewidth, 
$\left( \Gamma_{31} + \Gamma_{32} \right) /2$.

The above analysis is for a stationary atom. If the atom 
is moving, the laser frequency and detuning as seen by 
the atom are velocity dependent. To obtain the probe 
absorption in a gas of moving atoms, the above 
expressions have to be corrected for the velocity of the 
atom and then averaged over the 
Maxwell-Boltzmann distribution of velocities. Such an 
analysis has been done in Ref.\ \cite{VAR96}, and the 
important conclusion is that the location of the peaks 
given in Eq.\ (1) does not change, but the linewidths are 
now given by
\begin{equation}
\Gamma_{\pm} = \frac{\Gamma_{31} + \Gamma_{32} + 2D}{4} 
\left( 1 \mp \frac{\Delta_c}{\sqrt{\Delta_c^2 + 
\Omega_R^2}} \right) . 
\end{equation}
Here, $D$ is the usual Doppler width, which is 560 MHz 
for room-temperature Rb atoms. Thus, the earlier 
conclusions are still valid, except that the unperturbed 
linewidth is now $\left( \Gamma_{31} + \Gamma_{32} +2D 
\right) /2$, which includes a Doppler broadening term.

The main idea for our experiment is clear from Eq.\ (3). 
For $\Delta_c \gg \Omega_R$, the peak at $\Delta_c /2 + 
\frac{1}{2} \sqrt{\Delta_c^2 + \Omega_R^2}$ has 
significantly smaller linewidth than the unperturbed 
value. Indeed, this was proposed by Vemuri {\it et al.\ 
}\cite{VAR96} as a means of achieving sub-Doppler 
resolution when Doppler broadening dominates the 
linewidth. Sub-Doppler linewidths were subsequently 
observed by Zhu and Wasserlauf \cite{ZHW96} in a Rb vapor 
using an intense control beam from a Ti-sapphire laser. 
Our work extends this to sub-natural linewidths and 
requires only about 25 mW of control-laser power, which 
is easily available from diode lasers. In our second 
technique, the unperturbed linewidth is close to the 
natural linewidth and even smaller powers of $\sim$1 mW 
are enough to observe sub-natural linewidth.

The experimental set up is shown schematically in Fig.\ 
\ref{schematic}. The probe and control beams are obtained 
from two frequency-stabilized diode laser systems 
operating near the 780 nm $D_2$ line in Rb. The linewidth 
of the lasers after stabilization has been measured to be 
below 1 MHz. The two beams 
co-propagate through the cell with orthogonal 
polarizations. For the second set of experiments, a 
counter-propagating pump beam is generated from the probe 
laser using a beamsplitter. 

For the first set of experiments, we used a control beam 
with power of 25 mW, corresponding to a Rabi frequency of 
about 200 MHz. Its detuning, $\Delta_c$, was varied from 
0 to $-1.18$ GHz. Probe-absorption spectra at different 
values of $\Delta_c$ are shown in Fig.\ \ref{sub1}(a). 
The unperturbed probe absorption is Doppler broadened to 
560 MHz, as shown in the top trace. Therefore, the 
Doppler width dominates the linewidth. As expected from 
Eq.\ (1), there are two peaks in each spectrum. The first 
peak is near $\Delta = 0$, and its width is close to the 
Doppler width, while the second peak lies near $\Delta = 
\Delta_c$ and is much narrower. At small values of 
$\Delta_c$, the second peak lies within the Doppler 
profile of the first peak. According to Eq.\ (3), the 
width of the second peak decreases as the detuning is 
increased. This is indeed what we observe in Fig.\ 
\ref{sub1}(b) where the detuning is $-1180$ MHz. The peak 
lies well outside the Doppler profile of the first peak 
(not shown) and has a sub-natural linewidth of only 5.1 
MHz. However, a quantitative comparison of this linewidth 
to that in Eq.\ (3) is not justified because, as 
mentioned before, there are two levels in the excited 
state, $F'=1$ and 2, that can act as level $\left| 3 
\right>$ of the $\Lambda$ system. The control laser 
dresses both these levels and, since the levels are 
separated by 157 MHz, the value of $\Delta_c$ for each 
level is different. Thus, the probe absorption is a 
convolution of these two absorption profiles. This is 
evident from Fig.\ \ref{sub1}(a) where the lower traces 
have non-Gaussian lineshapes. At larger detunings, the two 
levels act as an effective single level, but the 
linewidth is still probably limited by the difference in 
detuning.

For the second set of experiments, we used a 
counter-propagating pump beam with a power of 100 $\mu$W, 
compared to the probe power of 10 $\mu$W. In this 
configuration, the zero-velocity group of atoms 
preferentially absorbs from the pump beam and the probe 
gets transmitted. This is a standard technique to 
overcome the first-order Doppler effect, and the probe 
transmission shows narrow peaks at the location of the 
excited-state hyperfine levels, as seen in the middle 
trace of Fig.\ \ref{sub2}(a). Ideally, the linewidth of 
the hyperfine peaks should be the natural linewidth, but 
our observed linewidth is increased to about 13 MHz. 
There are several effects that contribute to this 
increase; the most important are power broadening due to 
the pump beam and a small misalignment angle between the 
counter-propagating beams \cite{foot1}.

For the bottom trace in Fig.\ \ref{sub2}(a), the control 
laser is also turned on. The laser is detuned by +11.5 
MHz from the $5P_{3/2},F'=2$ level. It creates two 
dressed states near this level, and the $F'=2$ peak in 
the probe spectrum splits into an Autler-Townes doublet. 
In Fig.\ \ref{sub2}(b), we zoom into this region. The 
control-laser power is 3 mW, corresponding to a Rabi 
frequency of 25 MHz. The doublet peaks are well separated 
and show linewidths of 9.1 MHz and 3.7 MHz, respectively. 
As expected from Eq.\ (2), the sum of the two linewidths 
is equal to the unperturbed linewidth of $\sim$13 MHz. In 
fact, with a value of 12.8 MHz for the unperturbed 
linewidth, the widths of the two peaks calculated from 
Eq.\ (2) are exactly equal to the observed values. At 
lower powers (corresponding to smaller values of 
$\Omega_R$), the linewidth is still smaller and we have 
observed linewidths down to 2 MHz. However, the lineshape 
is not perfectly Lorentzian, partly because the peaks lie 
on the side of the Doppler profile. It is possible to 
remove the Doppler profile using several methods, such as 
by subtracting the absorption of a second identical probe 
beam, but we have not attempted this so far.

The configuration with the pump beam has the advantage 
that the control laser, because of its small detuning and 
Rabi frequency, only dresses one of the hyperfine levels 
in the excited state. Thus, this is closer to an ideal 
three-level system and the theoretical predictions can be 
applied with greater confidence. To test this, we have 
studied the separation of the two dressed states as a 
function of control-laser power at a fixed detuning of 
+11.5 MHz. The results are shown in Fig.\ \ref{peaksep}. 
The solid line is the predicted variation from Eq.\ (1) 
and the measured separation agrees very well with the 
prediction. The linewidths of the two peaks also follow 
the dependence given in Eq.\ (2), with the correction 
that the unperturbed linewidth is broadened to about 13 
MHz.

In conclusion, we have demonstrated that it is possible 
to get sub-natural linewidth for probe absorption through 
a room temperature atomic gas. The Doppler width is 
reduced by more than a factor of 100 using a control 
laser that drives the excited state on a second 
transition of the three-level $\Lambda$ system. The basic 
mechanism that modifies the probe absorption is the 
quantum coherences in the excited state created by the 
control laser. The creation of resonances with 
sub-natural linewidth may have immediate applications in 
precision spectroscopy and better stabilization of lasers 
on atomic transitions. It could also be used to achieve 
sub-Doppler temperatures in laser cooling of ions. Such 
low temperatures are important for future applications in 
quantum computing using trapped ions where the ion needs 
to be cooled to the quantum-mechanical ground state.

We thank Hrishikesh Kelkar for help with the 
measurements. This work was supported by a research grant 
from the Department of Science and Technology, Government 
of India.

\begin{figure}
\caption{
Three-level $\Lambda$ system in $^{87}$Rb. The control 
laser drives the $\left| 1 \right> \leftrightarrow \left| 
3 \right>$ transition with Rabi frequency $\Omega_R$ and 
detuning $\Delta_c$. The probe laser measures the 
absorption on the $\left| 1 \right> \rightarrow \left| 2 
\right>$ transition at a detuning $\Delta$. $\Gamma_{31}$ 
and $\Gamma_{32}$ are the spontaneous decay rates from 
the excited state.
}
\label{levels}
\end{figure}

\begin{figure}
\caption{
Schematic of the experiment. The probe and control beams 
are derived from diode laser systems. The power in each 
beam is set using a half-wave plate ($\lambda/2$) and a 
polarizing beamsplitter (PBS). The two beams are chosen 
to have orthogonal polarizations so that they can be 
mixed and separated using PBS's. The probe beam is 
detected on a silicon photodetector (PD). A part of the 
probe beam is split using a beamsplitter (BS) and used as 
a counter-propagating pump for the second set of 
experiments. The angle between the 
counter-propagating beams is close to 0 and has been 
exaggerated for clarity. BD's are beam dumps.
}
\label{schematic}
\end{figure}

\begin{figure}
\caption{
The figure shows the transmission of the probe beam as a 
function of frequency for various values of control-laser 
detuning, $\Delta_c$. In (a), the top trace is the 
unperturbed probe absorption showing the usual Doppler 
profile. The lower traces have a second narrow peak whose 
location is given by Eq.\ (1). The values of $\Delta_c$ 
are small enough that the second peak lies within the 
Doppler profile of the first. The control laser also 
affects the lineshape of the 
Doppler-broadened peak and makes it non-Gaussian. In (b), 
the control-laser detuning is increased to -1180 MHz. We 
zoom into this peak which lies away from the Doppler 
profile of the first peak. The dashed line is a 
Lorentzian fit and yields a full-width of 5.1 MHz, 
compared to the Doppler width of 560 MHz and the natural 
linewidth of 6.1 MHz.
}
\label{sub1}
\end{figure}

\begin{figure}
\caption{
In (a), the top trace is the Doppler-broadened profile of 
the probe beam when both control and pump beams are off. 
In the middle trace, the 
counter-propagating pump beam is turned on. The various 
hyperfine transitions (and spurious crossover resonances) 
are clearly resolved. In the bottom trace, the control 
laser is also turned on. Since the laser is tuned close 
to the $ F'=2$ level, the $F'=2$ peak splits into an 
Autler-Townes doublet. In (b), we zoom into the doublet. 
The control laser has a Rabi frequency of 25 MHz and 
detuning of 11.5 MHz. The dashed line is a Lorentzian fit 
to the two peaks, and yields a full-width of 9.1 MHz for 
the larger peak and only 3.7 MHz for the smaller peak.
}
\label{sub2}
\end{figure}

\begin{figure}
\caption{
The separation of the two dressed states is shown as a 
function of power in the control beam, for a fixed 
detuning of $+11.5$ MHz. From Eq.\ (1), the separation is 
given by $\sqrt{\Delta_c^2 + \Omega_R^2}$, where 
$\Omega_R^2$ is proportional to the power. The solid line 
is the predicted variation from Eq.\ (1) and shows 
excellent agreement.
}
\label{peaksep}
\end{figure}


\begin{references}
\bibitem{foot1}
Collisional broadening in the vapor cell (estimated to be 
$<$100 kHz) and the linewidth of the laser ($<$1 MHz) do 
not contribute signficantly.
\end{references}
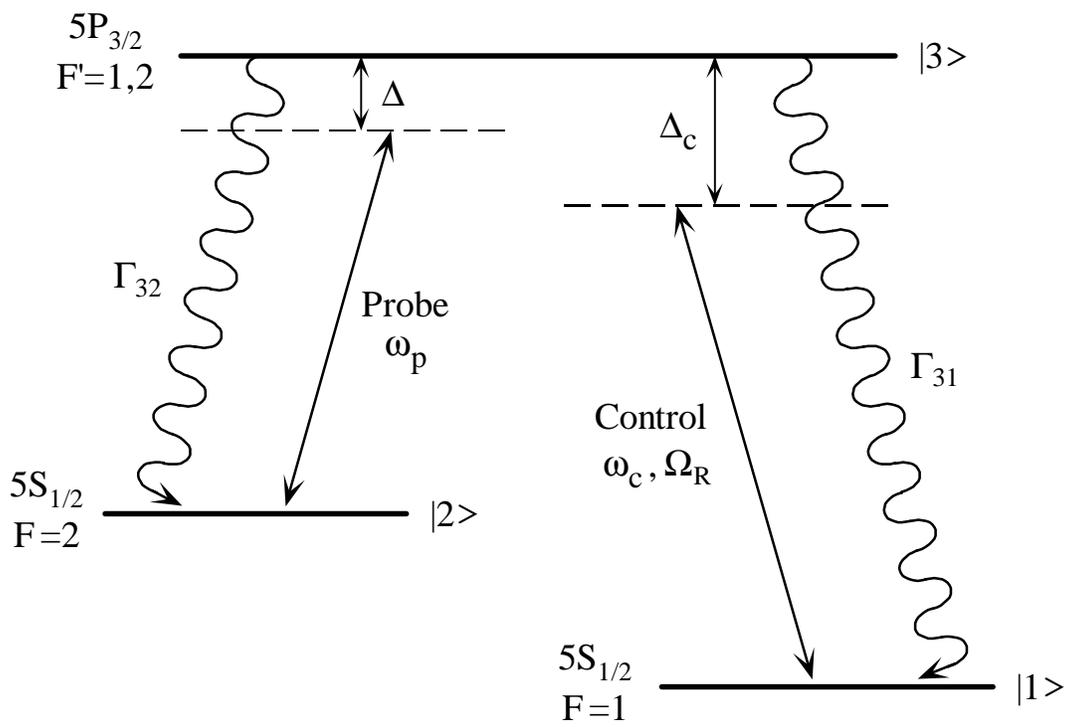
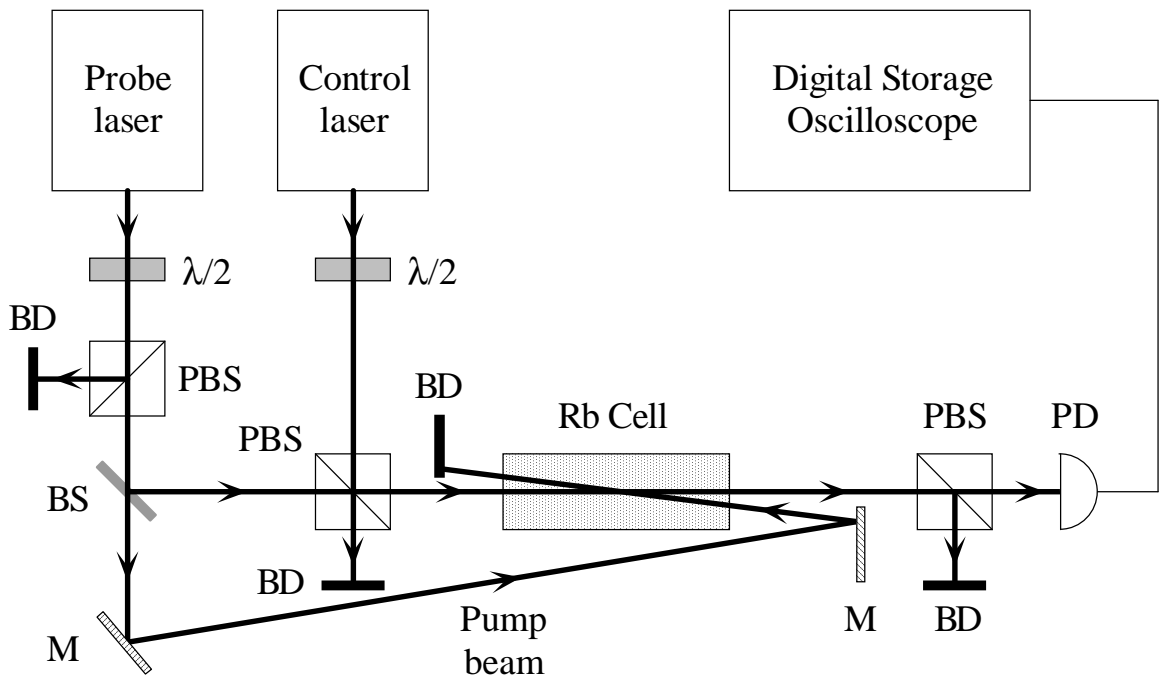
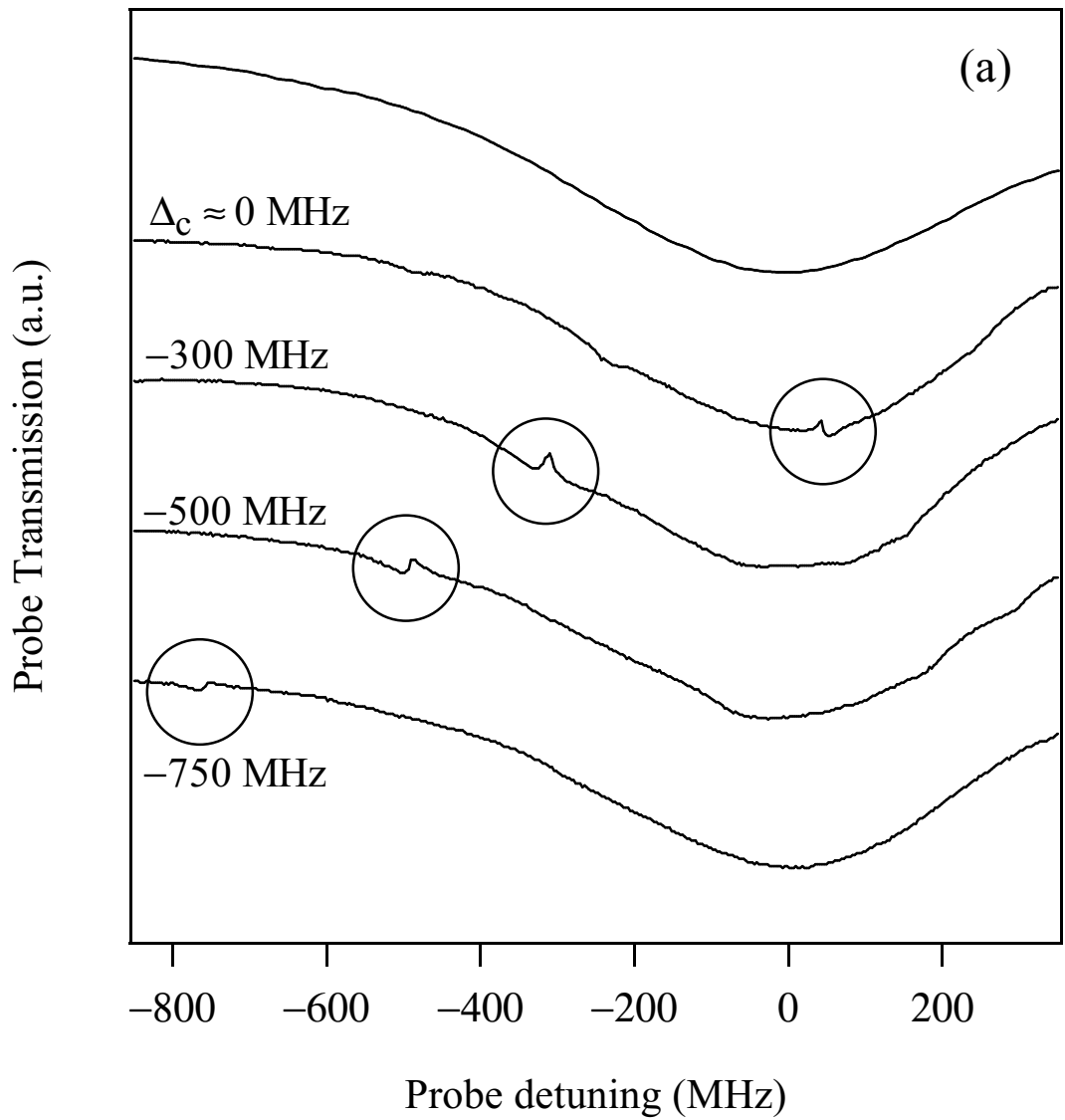
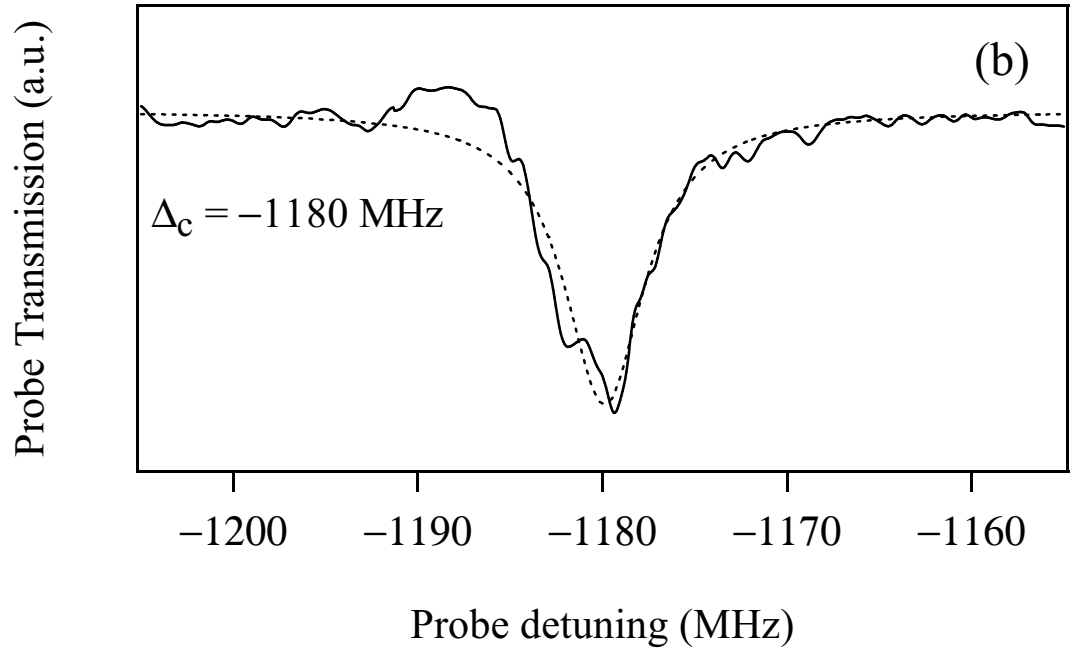
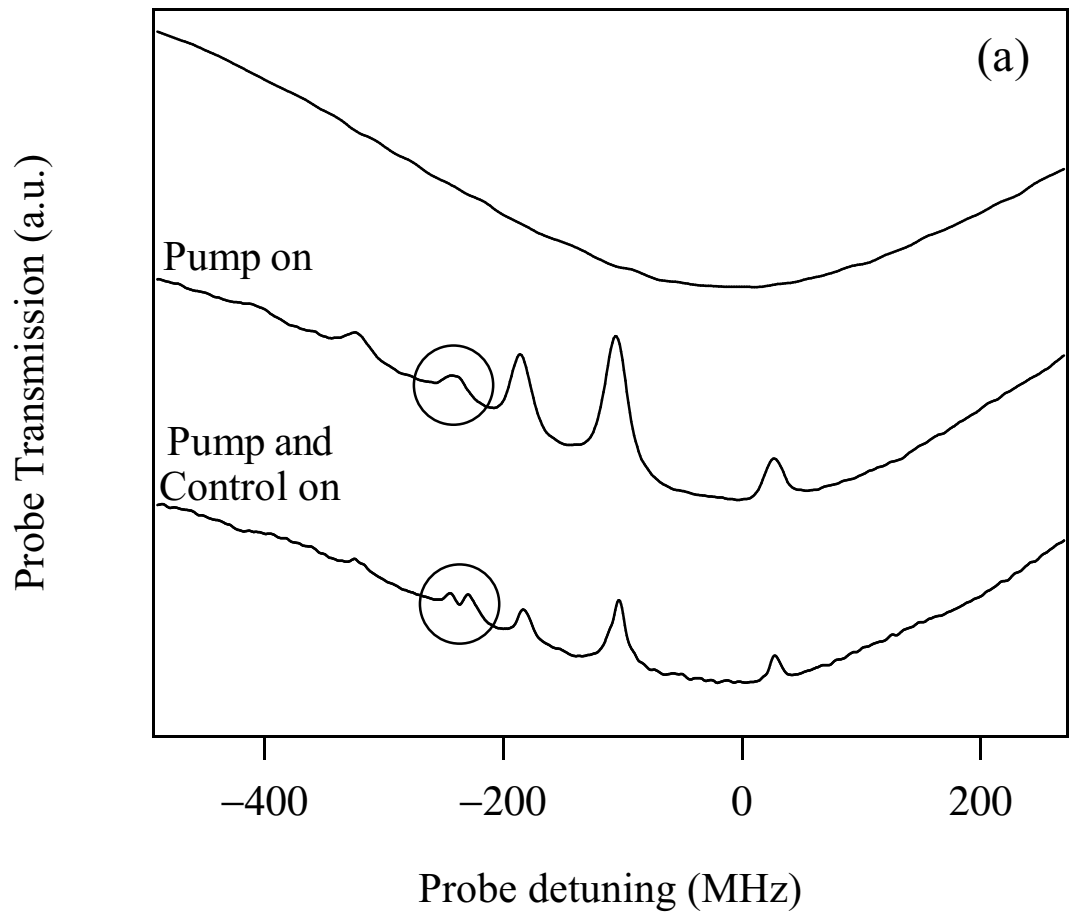
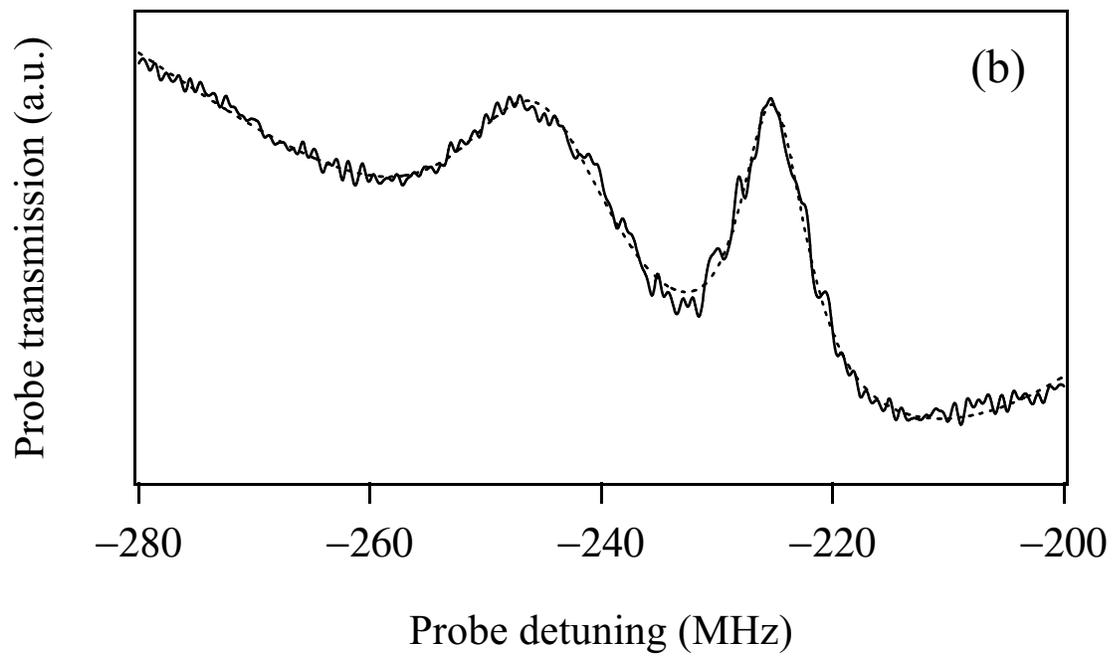
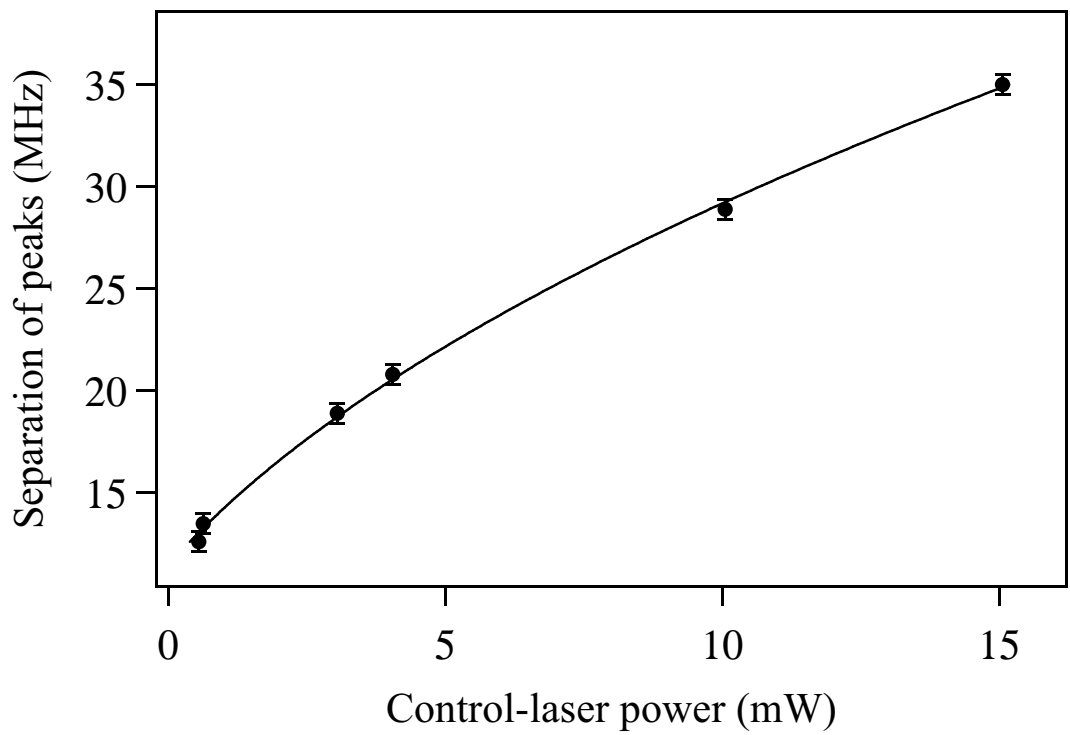
\end{document}